# Cryptocurrency Risk, Trust, and Acceptance in Thailand: A Comparative Study with Switzerland.

Kanyanut Suriyan
Siam Technology College, Bangkok Thailand
kunyanuts@siamtechno.ac.th

Tim Weingaertner
Lucerne University of Applied Sciences and Arts, Switzerland.
tim.weingaertner@hslu.ch

**Abstract:** The adoption of the "Pao Tang" digital wallet in Thailand, promoted under the "Khon la Krueng" (50-50 Co-Payment) Scheme, illustrates Thailand's receptiveness to digital financial instruments, amassing over 40 million users in just three years during the COVID-19 social distancing era. Nevertheless, acceptance of this platform does not confirm a broad understanding of cryptocurrencies and Web 3.0 technologies in the region. Through a mix of documentary research, online surveys and a targeted interview with the Pao Tang app's founder, this study evaluates the factors behind the Pao Tang platform's success and contrasts it with digital practices in Switzerland. Preliminary outcomes reveal a pronounced knowledge gap in Thailand regarding decentralized technologies. With regulatory frameworks for Web 3.0 and digital currencies still nascent, this research underscores the need for further exploration, serving as a blueprint for shaping strategies, policies, and awareness campaigns in both countries.

**Keywords**: Digital Wallets, Cryptocurrencies, DLT, Pao Tang, Risk, Regulatory Frameworks.

## 1. Introduction

### 1.1 Global Trend of Digital Financial Tools Adoption

The digital economy is continuously evolving, and digital wallet platforms have become pivotal tools for both individuals and businesses worldwide. These platforms not only serve as conduits for managing a myriad of digital assets, from loyalty points to cryptocurrencies, but also facilitate secure transactions. As illustrated in Figure 1, the adoption of such digital financial tools has been on the rise since 2020 and is forecasted to continue its growth trajectory till 2024. The global trend showcases a shift towards "cashless societies", propelled further by innovations in digital wallets, QR code transactions, and enhanced digital payment systems. In the context of this paper, we define a digital wallet as a secure mobile app or online service that allows individuals to store and manage various forms of digital assets and information, such as credit card details, bank account information, loyalty cards, or digital identities. While digital wallets can hold cryptocurrencies, they are not limited to them and can encompass a wide range of financial and identification tools and services.

### 1.2 "Pao Tang" in the Thai Digital Era

Thailand witnessed a significant digital shift during the COVID-19 pandemic, as Thai citizens rapidly adapted to the "Pao Tang" mobile application. With over 40 million users registered over the span of three years, this platform has become indispensable for daily transactions (Banchongduang, S., 2022). Furthermore, the "Pao Tang" app, developed by Krungthai Bank, supports various digital payment services. Initially conceived to back the government's social welfare and economic stimulus schemes during the pandemic, it's now integral to the country's digital financial landscape, bridging the digital divide, enhancing the financial system, and steering Thailand towards a cashless society (Krungthai Bank's Super App, 'Pao Tang', an All-in-one Platform for Thais, 2022).

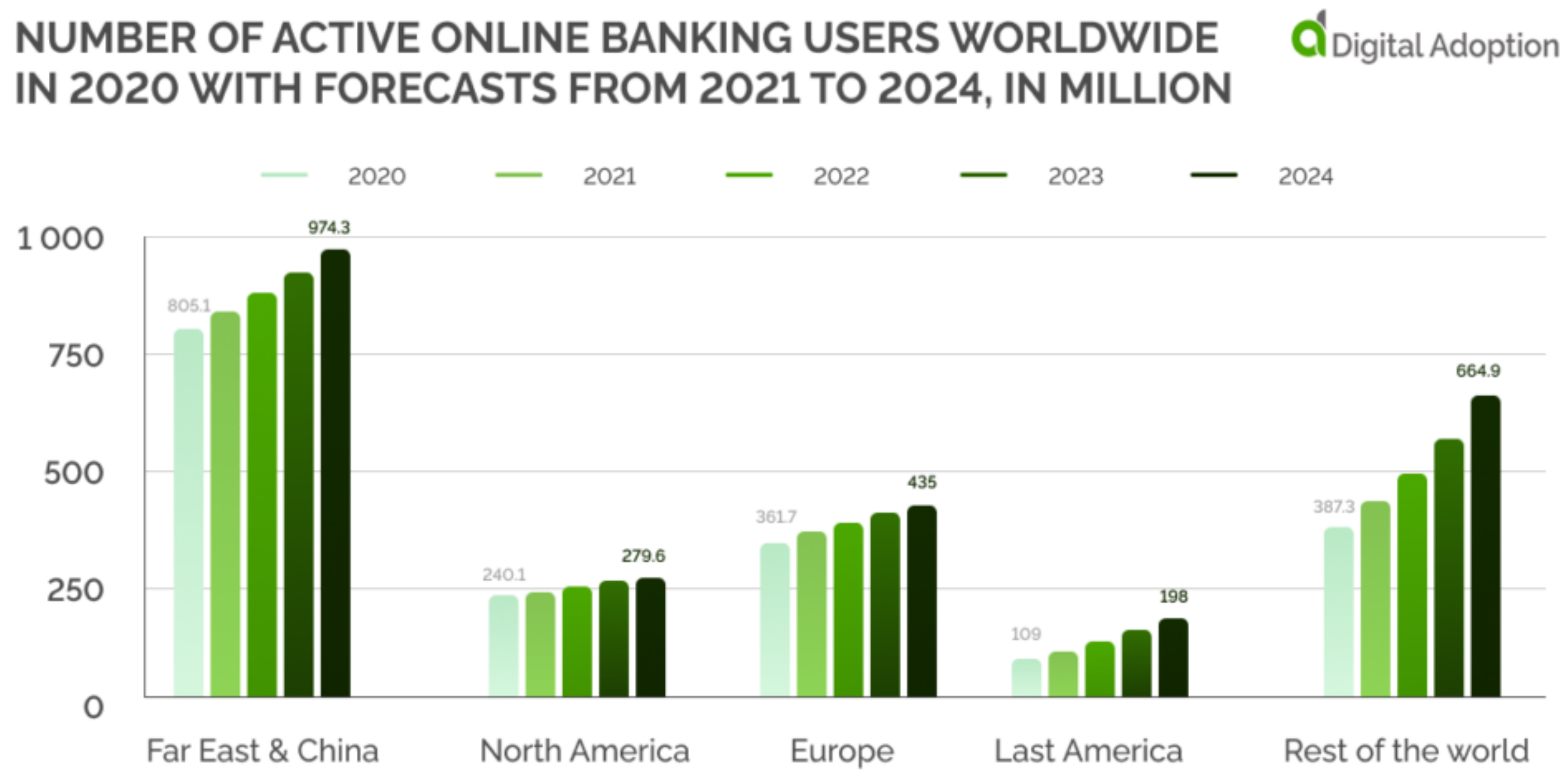

**Figure 1**: *Number of Active Online Banking Users Worldwide in 2020 with forecasts from 2021 to 2024, In million*. (Digital adoption available from Team, D. A. (2022))

### 1.3 Switzerland: A Comparative Study

Switzerland stands as an epitome of economic stability, with strengths in sectors ranging from banking to pharmaceuticals. The nation's esteemed financial sector, coupled with its emerging role as a pro-blockchain hub also called "Crypto Valley" in the canton of Zug is used frequently as a global benchmark and presents a fascinating contrast to Thailand's rapidly growing economy. While Thailand's economic strategies lean more towards protectionism, Switzerland champions liberal economic ideologies. Nevertheless, the percentage of Thai cryptocurrency account holders increases doubly from 2019 to 2022 compared with Swiss cryptocurrency account holders (see Figure 2). This research seeks to unravel how these contrasting nations navigate the realms of blockchain, providing insights into the regulation, usage, and risk awareness.

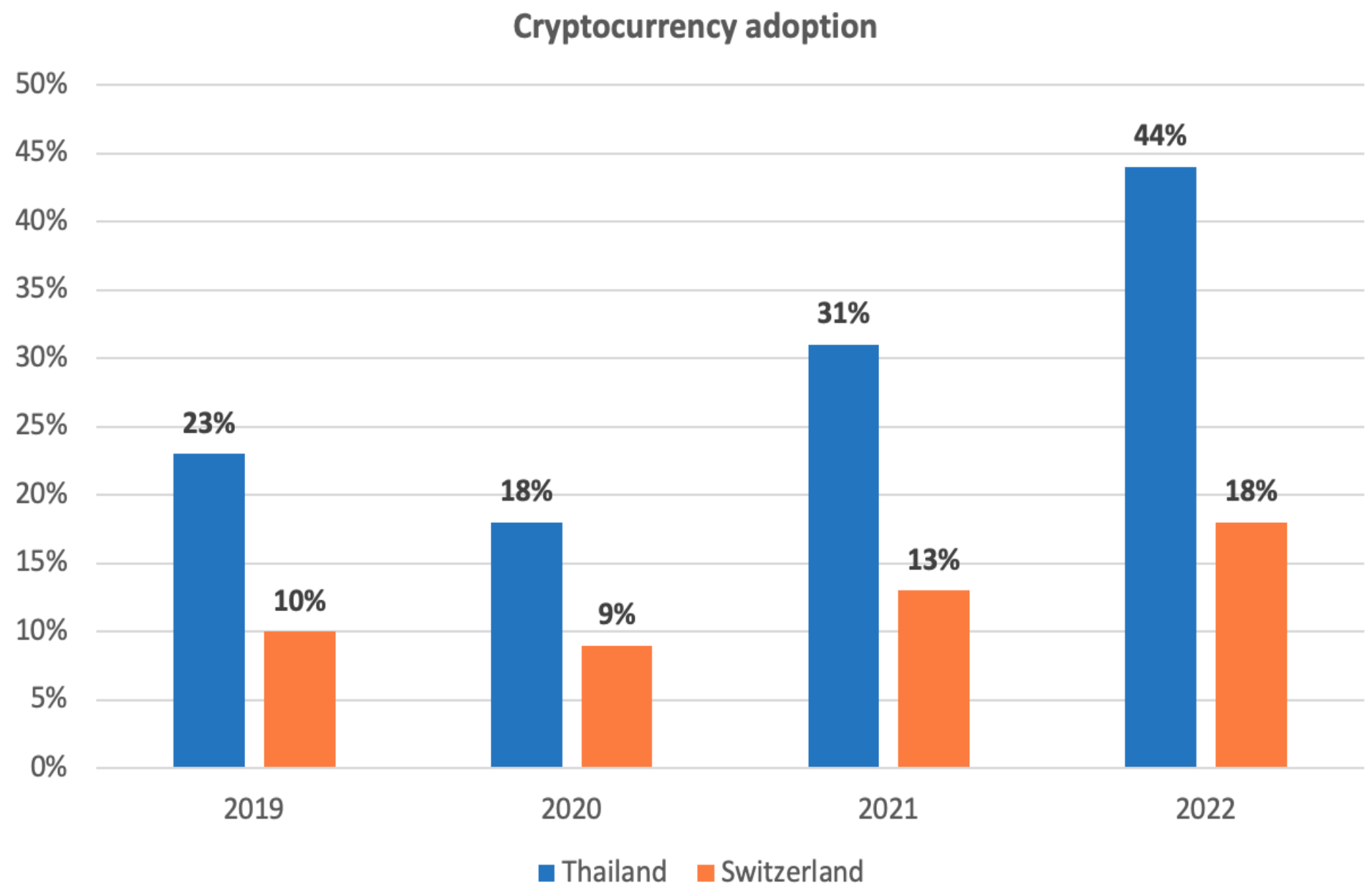

**Figure 2** *Cryptocurrency account holders. (Own diagram, data source Crypto ownership by country. (2023))*

To benchmark the trading volumes in cryptocurrencies between Thailand and Switzerland, we performed a comparison of trading volumes analogous to IFZ (2023). Figure 3 and 4 show results of this study.

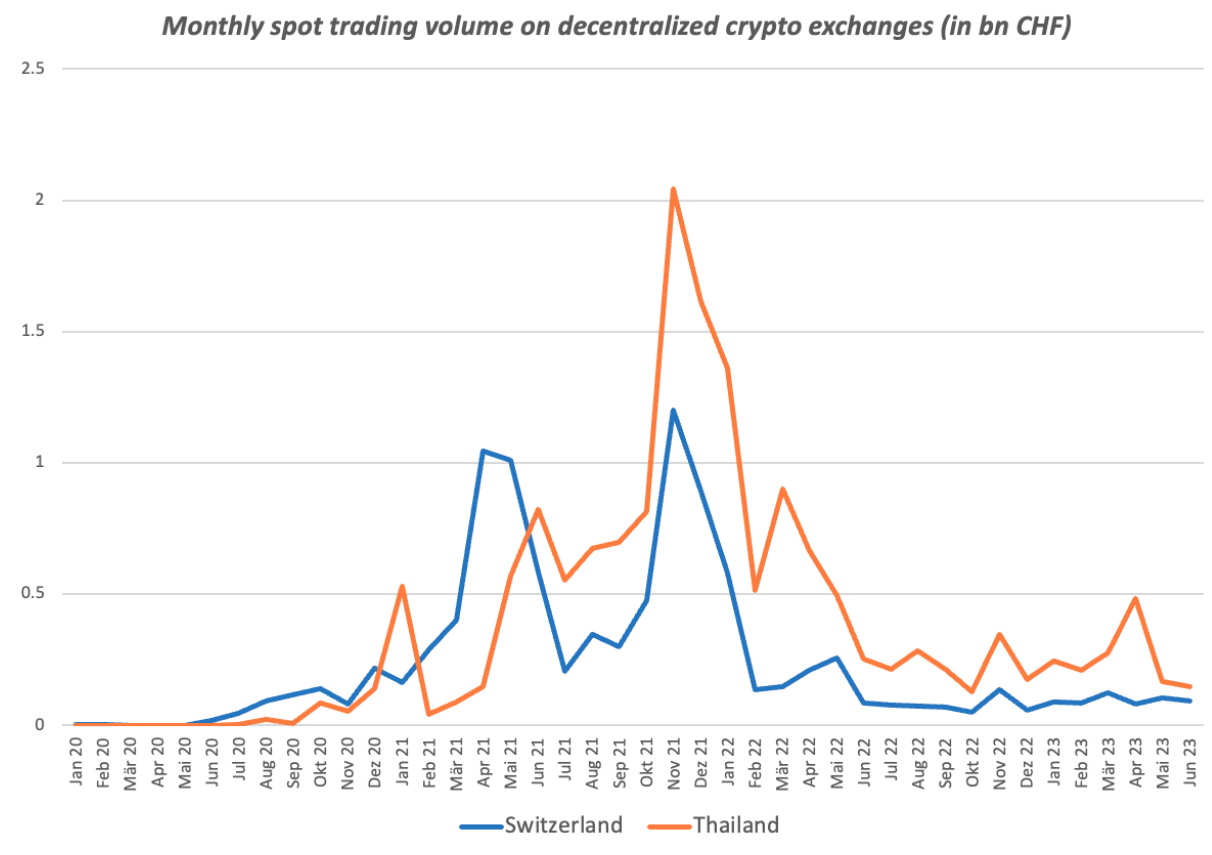

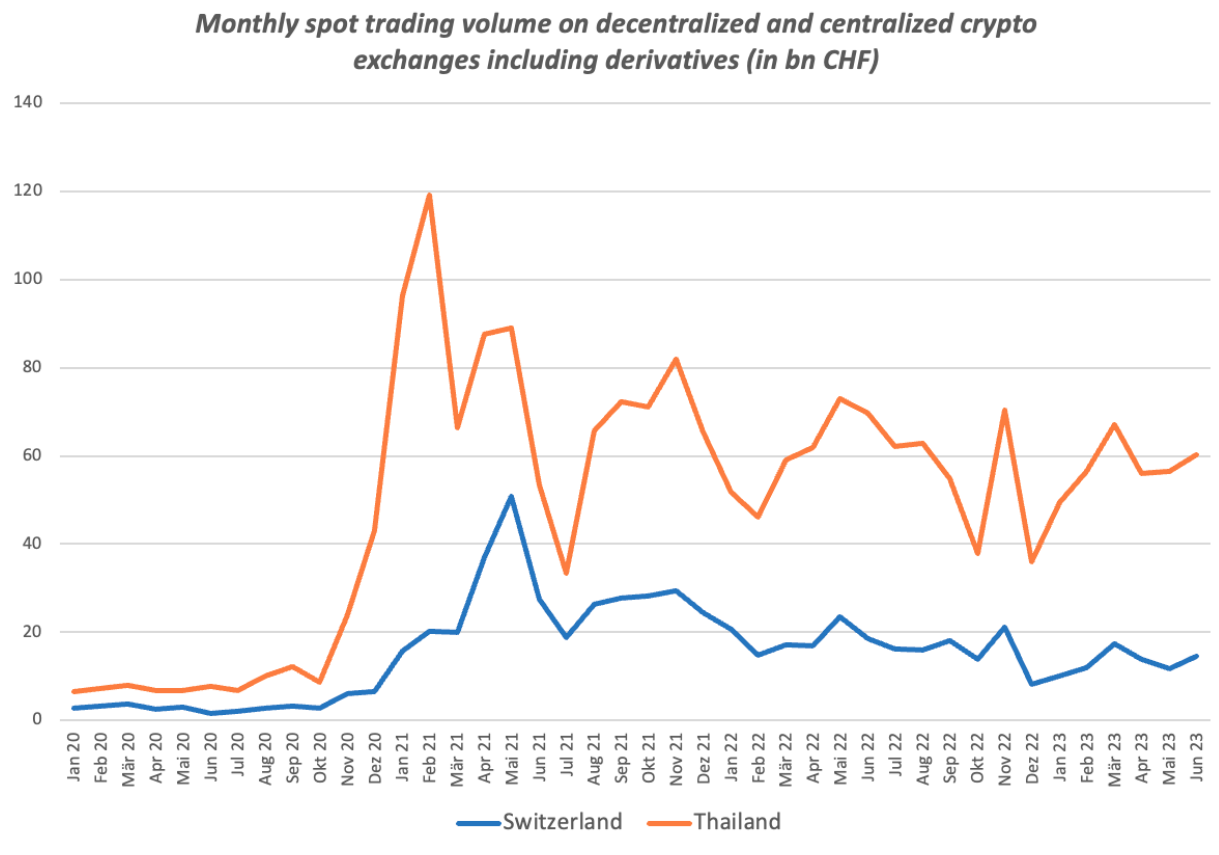

**Figure 3** *Monthly spot trading volume on decentralized crypto exchanges (in bn CHF, own diagram)*
**Figure 4** *Monthly spot trading volume on decentralized and centralized crypto exchanges including derivatives (in bn CHF, own diagram)*

### 1.4 Problem Statement

The digital transformation, marked by the surge of digital payments, brings to the fore several challenges and opportunities. While digital wallets, underpinned by decentralized technologies like blockchain and DLT, promise benefits such as increased financial autonomy, they also come with inherent risks. Despite the growing excitement around these wallets, a clear gap exists in users' understanding of the underlying decentralized technologies. This paper delves into this contrasting landscape. Main objectives include:

- Analyzing cryptocurrency acceptance and risk awareness in Thailand compared to Switzerland.
- Unraveling the factors behind the success of the "Pao Tang" app.
- Understanding the potential of integrating Web 3.0 technologies and cryptocurrencies in Thailand.

## 2. Current Financial Landscape and Recent Work

The financial landscape, driven by technology, reveals intricate details about a nation's economic posture, technological acceptance, and future trajectories. This section delves into understanding these nuances by studying the landscapes of both Thailand and Switzerland. Subsequent discussions offer insights into regulatory structures, comparative economic and cultural analysis, and reflections from prior academic work.

### 2.1 Digital Financial Landscape in Thailand

*Traditional vs. Digital Payments in Thailand:* In Thailand, traditional payment methods have been the norm for many years. However, technological advancements have caused a paradigm shift towards digital payment mechanisms, such as the QR-code payments and digital wallets. Such innovations are enhancing financial services and narrowing financial divides for both businesses and households (Financial Landscape for Digital and Sustainable Economy, n.d.).

*Challenges in the Digital Era:* Technological adoption is not without challenges. If businesses and households fail to adapt, there's potential for widening economic disparities, including issues like rising household debts. Thus, the Thai financial sector's challenge is to harmonize innovation with risk management (Moenjak et al., 2020).

*The Role of the Pao Tang App:* "Khon la Krueng" (Let's Go Halves/50-50 Co-Payment) scheme was one of the government's economic stimulus measures that help people, small business owners, hawkers, street vendors, and other businesses. Krungthai Bank has developed a registration website for people and shop owners. Eligible participants paid only 50% of the goods price and the government would transfer the other 50% of the price to merchants. Since payment and transfer of money must be made via the Pao Tang application, a product of Krungthai Bank, this app played a pivotal role in the scheme's success. "In the first phase of the Khon La Khrueng scheme, there were more than 15 million people registered and more than 1 million merchants participating, which generated more than 45,000-million-Baht circulating funds" (Krung Thai, 2022).

**2.2 Regulatory Framework in Thailand**

Thailand's approach to the rapidly emerging digital assets and cryptocurrencies has been proactive and comprehensive. The foundation for this approach is the regulatory directives established to govern the use, issuance, and trading of these assets.

The Royal Decree on Digital Asset Businesses B.E. 2561 (2018) decree forms the bedrock of Thailand's cryptocurrency regulation. It came into effect to provide clarity and to set clear parameters for both individuals and businesses engaged in digital asset transactions. Some key provisions include:

1. *Classification of Digital Assets:* The decree specifically categorizes digital assets into two primary types: Cryptocurrencies, serving as a medium of exchange, and Digital Tokens, which represent rights of a person in an investment or to acquire goods and services. This distinction aids in providing specific regulations for each type (Legal Counsel and Development Department, The Office of the Securities and Exchange Commission, 2018).
2. *Registration Requirements*: Entities wishing to engage in digital asset businesses are required to register with the Securities and Exchange Commission (SEC) within 90 days. This is vital for monitoring and supervisory purposes.
3. *Protection Measures*: To safeguard consumers and maintain market integrity, the Bank of Thailand has prohibited commercial banks from direct participation in cryptocurrency transactions. However, they have given green lights to certain companies as exchanges and dealers after a rigorous vetting process.

4. *Central Bank Digital Currency (CBDC):* The project "Inthanon" was introduced, representing the Bank of Thailand's interest in creating a CBDC to further streamline and secure financial transactions.(Legal Counsel and Development Department The Office of the Securities and Exchange Commission, 2018)
5. *Obligations for Digital Token Issuers:* Entities intending to publicly offer Digital Tokens must be either a limited company or public limited company established under Thai law. Before any public offering, they must obtain approval from the SEC. Furthermore, these entities have ongoing disclosure requirements, including updating the SEC about their financial status, business operations, and other pertinent information (Legal Counsel and Development Department, The Office of the Securities and Exchange Commission, 2018).
6. *Role of the SEC:* The SEC has been empowered not just to oversee but also to grant exemptions. This ability ensures flexibility in adapting to evolving market dynamics without stifling innovation.

| | **Digital asset exchange** | **Digital assets broker** | **Digital assets Dealer** |
|---|---|---|---|
| **Status** | Center or a network | Person | Person |
| **Purpose** | Purchasing, selling, or exchanging of digital assets | Being a broker or an agent for any person in the purchase, sale, or exchange of digital assets to other person | Purchasing, selling, or exchanging digital assets for his/her own account |
| **Process/Operation** | Matching or arranging the counterparty or providing the system | In consideration of a commission, fee, or other remuneration | Outside the digital -asset exchange |

Table 1. Digital asset business in Thailand (Source: Security Exchange Commission of Thailand (Tamphakdiphanit & Laokulrach 2020)

While Thailand's regulatory stance is clear and well-structured, it reflects an understanding of the complex nature of digital assets. This progressive regulatory environment not only fosters innovation but also ensures that risks are managed, striking a balance that many nations strive to achieve.

### 2.3 Payment Systems and Regulatory Aspects in Switzerland

Switzerland, with its robust banking infrastructure, continues to embrace both traditional (like cash) and modern payment methods. During the COVID-19 pandemic, the number of cash payments has been reduced (Digitalisation trends in the Swiss payment landscape. (n.d.)). Mobile payment solutions like Apple Pay, Samsung Pay, and Google Pay have gained popularity in Switzerland. In addition, the Swiss e-payment solution TWINT, a mobile payment solution developed by Swiss banks, has gained massive traction in the last few years. TWINT allows users to make payments, transfer money, and even make online purchases. Due to the ease of using it with QR codes, many small businesses use this payment method. Moreover, Switzerland's proactive approach towards cryptocurrency, highlighted by its "Crypto Valley", underscores its commitment to innovation in the blockchain sector. Bitcoin and other cryptocurrencies can be used for payments in various establishments, and there are also Bitcoin ATMs available.

Switzerland has been adept in evolving its regulations to match the pace of technological advancements. In relation to blockchain, Switzerland has implemented a "technology-neutral" approach, ensuring that regulations are flexible enough to accommodate advancements without stifling innovation. In contrast to other countries, Switzerland refrained from creating its own blockchain law and instead adapted several existing laws (State Secretariat for International Finance SIF. (n.d.)).

The Swiss Financial Market Supervisory Authority (FINMA) has provided guidelines on Initial Coin Offerings (ICOs) (Finma (2018)). They have been used as a blueprint for many other countries and describe a distinction into three token classes: payment token, utility token, and security token. On 2 November 2022, FINMA introduced its revised anti-money laundering ordinance (AMLO-FINMA) with new provisions targeting virtual currency transactions. The Swiss National Bank (SNB) has set criteria for DLT trading facilities to access the Swiss Interbank Clearing (SIC) payment system, but they must first obtain a FINMA license (Schärli et al. 2023).

## 2.5 Economic and Cultural Landscape: Thailand vs. Switzerland

Comparing the economic and cultural landscapes of Thailand and Switzerland provides a deep insight into the underlying factors that influence their respective approaches to digital currencies.

### *2.5.1 Economic Overview*

Thailand's economy is a fascinating blend of its agrarian roots and modern sectors like tourism, manufacturing, and digital services. The nation's GDP is heavily reliant on exports, including electronics, automotive goods, and agricultural products. On the other hand, Thailand adopts a more protectionist approach in its economic policies, often emphasizing state control in vital sectors.

*Switzerland* stands tall as an economic powerhouse, bolstered by its industrial and service sectors. With industries ranging from banking, pharmaceuticals, and premium manufacturing (like watches), the nation's GDP per capita ranks among the highest globally. Switzerland's liberal economic ideologies further promote substantial economic liberty, providing a fertile ground for innovations, including those in the digital currency realm.

While Thailand's average monthly salary remains at about 19'410.77 Baht (482.34 CHF), Swiss residents enjoy an almost 12-fold higher income at approximately 224'386.09 Baht (5,572.84 CHF) (Rankings by Country of Average Monthly Net Salary, n.d.).

### *2.5.2 Cultural Dynamics*

The heartbeat of *Thailand* is its rich Buddhist heritage. Strong communal ties underscore its collectivist ethos. Traditional arts and crafts remain an integral part of daily life, reflecting the nation's profound cultural roots.

*Switzerland* presents a melting pot of diverse cultures, thanks to its linguistic diversity, including German, French, Italian, and Romansh. The Swiss value individualism, punctuality, and precision, which manifests in their work ethics and societal norms. Structured education and a strong leaning towards classical music and art form the foundation of Switzerland's cultural dynamics.

The economic compasses of Thailand and Switzerland point in distinct directions. While Thailand finds strength in sectors like agriculture and tourism, Switzerland emerges as a beacon in the financial realm. Such differences extend to their respective regulatory postures towards cryptocurrencies, with Switzerland possibly having a more encompassing risk assessment strategy.

## 2.6 Previous Studies on Digital Payments

The process of adopting and comprehending digital financial tools is steered by multiple factors. A detailed examination of prior studies furnishes a holistic view of these influential elements, especially emphasizing trust, perceived risks, and user acceptance.

Trust emerges as a pivotal determinant in the adoption of digital financial tools. Decaro & Saleh (2003) emphasized the essence of trust over technological advancements in online banking adoption. Echoing these sentiments, Patil et al. (2018) delved into the role of trust in mobile payments. They underlined its influential role in shaping both the behavioral intentions and the satisfaction levels of users. Furthermore, Bashir & Madhavaiah (2015) augmented the technology acceptance model by integrating trust, revealing that it significantly propels positive behavioral intentions in internet banking.

On the other hand, the perception of risk, especially in the digital sphere, profoundly impacts user acceptance. Liu et al. (2008) embarked on an exploration of internet banking user acceptance in contexts fraught with risk and uncertainty. Their conclusions emphasized the profound influence of perceived risk and uncertainty on acceptance levels. Supporting this notion, Decaro & Saleh (2003) contended that perceived risks and trust are inextricably linked when it comes to the adoption trajectory of online banking.

Delving into user acceptance, insights into the mechanisms and reasons behind the acceptance of new digital financial tools offer a comprehensive understanding of the broader dynamics at play in the digital economy. Pertaining to the context of Thailand, several scholars have imparted valuable insights. Lamsam et al. (2018) spotlighted the coexistence of traditional cash systems and electronic payments in Thailand, suggesting ways to optimize cash management efficiency. Achyar et al. (2022) identified international tourism receipts as a potent catalyst accelerating digital payment utilization in the country. Moreover, Khiaonarong (2000) provided a thorough examination of the evolution of electronic payment systems in Thailand, attributing a pivotal role to the central bank in the management and investment of these systems. Lastly, Gohwong (2017) categorized various digital payment forms in Thailand, singling out e-Money as the most prominent, closely followed by in-house funds transfer and payment cards.

## 3. Research Questions and Methodology

The main objectives of our research as described in section 1.4 first is to delve deep into the cryptocurrency acceptance and risk consciousness in Thailand compared to Switzerland, and second, to identify the factors that have made the Pao Tang wallet so popular, while also exploring the challenges and prospects of adopting cryptocurrencies and Web 3.0 technologies in Thai digital payments. This leads to the following research questions and their sub-questions:

**RQ 1:** How is the acceptance and risk awareness of cryptocurrencies in Thailand compared to that of Switzerland?

**RQ 1.1:** How is the acceptance of cryptocurrencies perceived among individuals and institutions in Thailand?

**RQ 1.2:** What is the level of risk awareness associated with cryptocurrencies among Thai stakeholders?

**RQ 1.3:** How is the acceptance and risk awareness of cryptocurrencies perceived in Switzerland?

**RQ 2:** What are the key success factors behind the Pao Tang App, and what challenges and opportunities arise from integrating cryptocurrency and Web 3.0 technologies in Thailand?

**RQ 2.1:** What are the key success factors behind the Pao Tang App in Thailand?

**RQ 2.2:** What opportunities exist for integrating cryptocurrency and Web 3.0 technologies within the Pao Tang App or similar platforms in Thailand?

To ensure comprehensive and sound research, different research methods were used. In the following, these are assigned to the research questions.

RQ 1.1 was addressed by a survey among Thai individuals, inquiring about their knowledge, attitudes, and acceptance of cryptocurrencies. The above survey was also used for RQ 1.2, focusing on questions that assessed the depth of their risk understanding related to cryptocurrency investments and transactions. RQ 1.3 was addressed by a comparative analysis of the same survey conducted in Switzerland. The results led to an answer of the overall RQ 1.

For RQ 2 a qualitative approach has been chosen. For RQ 2.1 and RQ 2.2 an interview with the Pao Tang app's founder Somkid Jiranuntarat, an adviser to the president of Krungthai Bank, which oversees the Pao Tang app and offered insights into the success factors, strategies, and user satisfaction metrics.

## 4. Results

This section presents a comprehensive analysis, shedding light on factors contributing to the success of the Pao Tang platform and offering a comparative exploration of user knowledge and trust in cryptocurrency and Web 3.0 technologies between Thailand and Switzerland.

### 4.1 Survey of User Knowledge and Trust in Cryptocurrencies in Thailand

In our endeavor to comprehend the underlying factors influencing the adoption and perception of cryptocurrencies and Web 3.0 technologies, a comprehensive survey was administered to respondents in Thailand (n=465). This survey aimed to gather quantifiable data on various demographics, unraveling correlations between age, gender, educational background, and their subsequent influence on cryptocurrency understanding, trust, and usage. By decoding these patterns these findings and inferences derived from this survey data:

- *Age:* In summary, older respondents in this dataset tend to find it more challenging to understand and use cryptocurrencies, feel they lack enough financial knowledge to understand them and are less aware that they can exchange cryptocurrencies with traditional currencies (Understand and use cryptocurrencies: 20-30 years: 43.43% don't use, 56.57% use; 31-40 years: 65.93% don't use, 34.07% use; 41-50 years: 62.79% don't use, 37.21% use; more than 50 years: 64.71% don't use, 35.29% use). Also older respondents tend to be less confident about the security aspects of cryptocurrency (20-30 years: 46.80% not confident, 53.20% confident; 31-40 years: 78.02% not confident, 21.98% confident; 41-50 years: 65.12% not confident, 34.88% confident; more than 50 years: 70.59% not confident, 29.41% confident)
- *Gender:* Male respondents are more likely to be aware of the advantages and disadvantages of cryptocurrencies (female: 41.83% aware, 58.17% unaware; male: 56% aware, 44% unaware). They are also more likely to follow news about cryptocurrencies regularly (female: 27.12% follow, 72.88% don't follow; male: 34.67% follow, 65.33% don't follow).
- There is no correlation between the level of education and usage of cryptocurrency.
- There is a statistically significant correlation between age and the influence of others suggesting that younger respondents are more likely to be influenced by others to use cryptocurrencies (20-30 years: 59.26% not influenced, 40.74% influenced; 31-40 years: 81.32% not influenced, 18.68% influenced; 41-50 years: 76.74% not influenced, 23.26% influenced; more than 50 years: 73.53% not influenced, 26.47% influenced).
- Among the people who view cryptocurrencies as high-risk (348), 162 people intend to or currently use cryptocurrencies for payments and 198 people intend to or currently use cryptocurrencies for speculation or financial gain. From this analysis, we can infer that even if they perceive cryptocurrencies as high-risk, a significant number of respondents still use or intend to use them.
- Among the respondents who find it convenient to use cryptocurrency anytime and anywhere (265), only 159 of them intend to or currently use cryptocurrencies for payments.

- There is a positive correlation between the belief that using cryptocurrencies can enhance one's financial goals and living standards, and the use of cryptocurrencies for speculation (0.418). This positive value indicates that respondents who believe that cryptocurrencies can improve their financial situation and standard of living are more likely to use cryptocurrencies for speculation.
- There is a higher correlation between those who are more aware of the advantages and disadvantages of cryptocurrencies and the usage for speculation (68.81%). This implies that awareness plays a more significant role in influencing the use of cryptocurrencies for speculation than for payments.

### 4.2 Comparative Survey in Switzerland

To draw a comparative insight, an analogous survey was conducted among respondents in Switzerland (n=79). Deploying a parallel set of questions, this survey aimed to uncover the intricacies of cryptocurrency and Web 3.0 technology adoption within a Swiss context. The subsequent table 2 provides a detailed account of the Swiss responses and how they contrast or align with the findings from Thailand.

| **Question** | **Thailand (Yes %)** | **Switzerland (Yes %)** | **Absolute Difference (Yes %)** |
|---|---|---|---|
| Are you aware of the advantages and disadvantages of Bitcoin or cryptocurrency investments | 46.88% | 92.41% | 45.52% |
| Do you actively follow cryptocurrency news? | 29.68% | 55.70% | 26.02% |
| Would you consider using cryptocurrencies as not requiring significant mental effort? | 72.26% | 26.58% | 45.68% |
| Would you consider it convenient to use cryptocurrency anytime and anywhere? | 56.99% | 67.09% | 10.10% |
| Do you possess the necessary resources and understanding to effectively use cryptocurrencies? | 43.44% | 65.82% | 22.38% |
| Do you view cryptocurrencies as a high-risk investment? | 74.84% | 91.14% | 16.30% |

| Question | Thailand (Yes %) | Switzerland (Yes %) | Absolute Difference (Yes %) |
|---|---|---|---|
| Would you say you possess sufficient financial knowledge to understand cryptocurrencies? | 40.65% | 77.22% | 36.57% |
| Do you know that you can exchange cryptocurrencies with Swiss francs or other currencies like any traditional money? | 37.57% | 97.47% | 60.05% |
| Are you confident about the security aspects of cryptocurrency? | 43.66% | 55.70% | 12.04% |
| Do you intend to or currently use cryptocurrencies for payments? | 41.29% | 30.38% | 10.92% |

Table 2. Comparison of both surveys in Thailand and Switzerland, listing the most significant differences between both countries (difference > 10%).

### 4.3 Analysis of the Pao Tang Platform's Success

Derived from the interview with Somkid Jiranuntarat, the Pao Tang app amplified its reach through multiple government-backed projects, notably "Chip-Shop – Chai" and "Khon la Krueng". These initiatives aimed to foster a culture of digital payment adoption among the Thai populace. Several determinants underscored its success:

- *Scalable Architecture:* The app's infrastructural design ensured it could manage a burgeoning user base without compromising efficiency.
- *User Experience:* Emphasizing intuitiveness, the app's interface was designed to cater to users spanning various age brackets, including the elderly.
- *Governmental Backing:* State support, especially through the mentioned campaigns, provided the requisite thrust to the app's adoption and incentivized citizens to use the app.
- *Cost Efficiency:* The platform's operations streamlined services, driving down costs while simultaneously widening its outreach.
- *Versatility:* Beyond mere payments, Pao Tang diversified by integrating a suite of services, like lottery, into its framework.
- *Future Vision:* The app's envisioned trajectory moves from a basic wallet, escalating to a more complex digital ecosystem, envisaging an open platform allowing for holistic third-party integrations.

However, despite its digital-forward stance, the platform currently has no roadmap to incorporate cryptocurrencies or blockchain integrations. Somkid Jiranuntarat recognized the advantages of cryptocurrencies, especially in fostering innovation and tokenization. Yet, the challenges were equally pronounced, encompassing the need for comprehensive public education, potential misuse, speculative tendencies, and the nuances of regulating digital assets versus digital currencies. Drawing inspiration from countries like Switzerland could pave the way forward for Thailand. Initiatives like the Eastern Economic Corridor (EEC) could provide a conducive environment for testing these technologies, leveraging a sandbox approach. A recurrent challenge echoed the scarcity of skilled personnel in emerging technologies within Thailand.

## 5. Discussion

In this section, we delve into the findings derived from the survey conducted among respondents in Thailand and Switzerland and the interview with Pao Tang's founder Somkid Jiranuntarat. The analysis unveils the levels of acceptance, understanding, and risk awareness associated with cryptocurrencies in both countries.

**RQ 1.1:** *How is the acceptance of cryptocurrencies perceived among individuals and institutions in Thailand?*

The data from Thailand illuminates a growing acceptance of cryptocurrencies, especially among younger and male demographics. Interestingly, despite the high-risk perception, there is a pronounced inclination towards both speculative engagements and genuine payment uses. The fear surrounding cryptocurrency security, more palpable among older respondents, underscores the necessity for bolstered public education on this frontier.

**RQ 1.2:** *What is the level of risk awareness associated with cryptocurrencies among Thai stakeholders?*

A significant 74.84% of Thai respondents categorize cryptocurrencies as high-risk. Yet, this apprehension does not deter them from either speculative engagements or direct payments. This suggests a complex interplay between risk perception and the perceived benefits or necessities driving cryptocurrency engagement among Thai stakeholders.

**RQ 1.3:** *How is the acceptance and risk awareness of cryptocurrencies perceived in Switzerland?*

Swiss data showcases a high level of cryptocurrency acceptance, coupled with comprehensive risk awareness. Notably, while 91.14% recognize the inherent risks, a substantial proportion (59.49%) still ventures into speculative cryptocurrency engagements. This suggests that Swiss stakeholders are not only well-informed but are also willing to engage with cryptocurrencies despite the associated risks.

This leads to the following findings for RQ 1: *How is the acceptance and risk awareness of cryptocurrencies in Thailand compared to that in Switzerland?*

*Awareness and Knowledge:* A significant majority of respondents from Switzerland (92.41%) are aware of the advantages and disadvantages of Bitcoin or cryptocurrency investments compared to about half (46.88%) from Thailand. Similarly, 77.22% of Swiss respondents claim to have sufficient financial knowledge to understand cryptocurrencies, in contrast to only 40.65% of Thai respondents. This suggests that Swiss respondents, on average, feel more knowledgeable about cryptocurrencies than their Thai counterparts. Such disparities may be rooted in cultural and educational differences, with Switzerland's rich financial heritage potentially fostering a deeper literacy.

*Cryptocurrency as a High-Risk Investment:* Datasets from both countries reflect a belief that cryptocurrencies are high-risk investments. However, this sentiment is even more pronounced among Swiss respondents (91.14%) compared to Thai respondents (74.84%). This variance may reflect a deeper understanding or exposure to financial matters, possibly driven by Switzerland's robust financial sector. Moreover, Switzerland's longstanding tradition of financial literacy could contribute to more informed perspectives on the inherent volatility and risk associated with cryptocurrencies. Conversely, the lower risk awareness in Thailand may indicate a need for enhanced financial education and outreach to foster a more nuanced understanding of cryptocurrency risks.

*Cryptocurrency Usage and Speculation:* The intention to use or currently use cryptocurrencies for speculation or financial gain is slightly higher among Swiss respondents (59.49%) compared to Thai respondents (48.82%). The convenience of using cryptocurrency anytime and anywhere is perceived higher among Swiss respondents (67.09%) than Thai respondents (56.99%). On the other hand, there is a slight preference among Thai respondents for using cryptocurrencies as a payment method (41.29%) over speculation or financial gais. This suggests a potential cultural or regulatory divergence in cryptocurrency utilization. Switzerland’s established financial infrastructure might foster a speculative approach, while in Thailand, a growing acceptance of digital transactions might be encouraging the use of cryptocurrencies for everyday payments. These trends reflect the evolving narratives of cryptocurrencies within different socio-economic contexts.

*Cryptocurrency Security and Resources:* More Swiss respondents (65.82%) feel they possess the necessary resources and understanding to use cryptocurrencies effectively compared to Thai respondents (43.44%). Confidence in the security aspects of cryptocurrency is slightly higher in the Swiss dataset (55.70%) compared to the Thai dataset (43.66%).

*Influence of Peer Group and Community:* The influence of the personal environment is slightly higher in Thailand (34.84% vs. 26.58% in Switzerland). In Thailand the age plays a statistically significant role while in Switzerland the age has no significant influence. This divergence might be indicative of deeper societal structures and cultural predilections influencing financial behaviors.

In summary, Swiss respondents, based on the conducted survey, generally seem more informed and more aware of the risks that come with the usage of cryptocurrencies compared to Thai respondents. This could be influenced by numerous factors, including financial infrastructure, education, cultural attitudes towards investments, and exposure to global financial trends.

**RQ 2.1:** *What are the key success factors behind the Pao Tang App in Thailand?*

The key success factors behind the Pao Tang app in Thailand include its scaling architecture, outstanding usability, and government support through initiatives like "Chip-Shop – Chai" and "Khon la Krueng" that encourage digital payment adoption. The app's ease of use, QR code-based payment system, and its adoption by small businesses and elderly individuals significantly contribute to its success. Moreover, Pao Tang has reduced service costs and enhanced accessibility to a broader population segment, gradually evolving from a wallet app to envisioning an open digital ecosystem, amplifying its usability and scope.

**RQ 2.2:** *What opportunities and challenges exist for integrating cryptocurrency and Web 3.0 technologies within the Pao Tang App or similar platforms in Thailand?*

Opportunities for integrating cryptocurrency and Web 3.0 technologies within the Pao Tang app or similar platforms in Thailand include driving innovation, fostering an open economy, and enabling tokenization and digitization of physical assets. Challenges encompass the need for public education on these technologies, risks of misuse and scams, speculative behaviors, and the regulation of digital assets versus digital currencies. Moreover, the demand for skilled individuals in these technologies, who are currently scarce, presents a significant hurdle. The Eastern Economic Corridor (EEC) is suggested as a potential testbed for such technologies, reflecting a sandbox approach to foster learning from other countries' experiences like Switzerland.

## 6. Conclusion and Recommendations

Based on our research we can state that in both Thailand and Switzerland there is a functioning digital payment ecosystem. The need for cryptocurrencies in payment is limited and a slightly larger group is using cryptocurrencies for speculation. Most people in Thailand are not aware of the risks, which gives the potential for further education.

Nevertheless, cryptocurrencies offer additional chances, like decentralization, cross border payments, financial inclusion, low fees, open ecosystems, peer-to-peer transactions, and a high innovative potential. Especially for Web 3, global accessibility, ownership over tokens and NFT, and smart contract secured transactions are major advantages.

From our research we derive the following recommendations:

- *Combination of centralized and decentralized payment system:* Web 3 offers a new and global possibility for many people and economies. While centralized digital payment systems are proven and functioning in many countries, a combination with decentralized, blockchain based cryptocurrencies and tokens enfold the full potential of Web 3 use cases. In order to benefit from this innovation, further research like Istrefaj, A. (2023) and sandboxes could help to take advantage of this opportunity.
- *Educational Workshops and Seminars:* One of the primary barriers to understanding decentralized technologies is a lack of foundational knowledge. Only 50.74% of Thai respondents are aware of the advantages and disadvantages of Bitcoin or cryptocurrency investments (92.41% in Switzerland). This disparity in perceived understanding suggests that there is room for educational improvement in Thailand. Partner with universities, colleges, and tech institutions in Thailand to host workshops, seminars, and courses dedicated to blockchain and decentralized technologies.
- *Regulatory Collaboration and Public Campaigns:* The regulatory frameworks between Thailand and Switzerland differ. There is still uncertainty with the industry and public. Thai regulatory bodies should establish clear guidelines and standards for digital wallet operations that can create a safer and more trustworthy environment for users.

- *Grassroots Community Building:* Another possibility is the usage of communities since they play a pivotal role in fostering learning and sharing experiences. Encouraging grassroots communities in Thailand can help disseminate such fundamental information more effectively. By encouraging peer-to-peer learning, early adopters and enthusiasts can share their experiences and knowledge with newcomers. The Eastern Economic Corridor (EEC)[1] offers a chance to build a sandbox and testbed for experimenting with new regulations and approaches to this technology.
- *Clear Licensing and Registration:* Only 45.22% of Thai respondents are confident about the security aspects of cryptocurrency. A clear licensing system can enhance trust and security confidence among users by ensuring that platforms adhere to standardized security practices. This would help ensure that only legitimate and compliant entities operate in the market.
- *Consumer Protection Mechanisms:* With 78.31% of Thai respondents viewing cryptocurrencies as high-risk investments, robust consumer protection mechanisms can address these concerns and build confidence in the technology. Establishing mechanisms to protect consumers from fraud, market manipulation, and platform insolvencies could support the trust in this technology. This could be achieved by including mandatory insurance for exchanges or dedicated funds to compensate users in case of losses.

## 8. Acknowledgments

We extend our heartfelt gratitude to all the survey participants who generously gave their time and shared their insights, enabling us to gather invaluable data for this research. Special thanks are due to our interview partner, Somkid Jiranuntarat, the innovator of the Pao Tang app and the contributor to our perspectives which enriched our understanding and significantly contributed to the depth of this study. We are also very thankful to the President of Siam Technology College, Asst. Prof. Pornpisud Mongkhonvanit, for his guidance and for providing the opportunity to undertake this research study. Lastly, our appreciation goes to Levin Reichmuth and Thomas Ankenbrand from IFZ for the data delivery of the comparative study in Figure 3 and 4.

[1] https://www.eeco.or.th/en

## 9. References

Achyar, D. H., Hasyyati, Z., Yumni, H., & Wafda, F. (2022, March). Acceleration of international tourism improves digital payments usage: The case of Thailand. In 2022 International Conference on Decision Aid Sciences and Applications (DASA) (pp. 212-214). IEEE.

Al-Dmour, A., Al-Dmour, H. H., Brghuthi, R., & Al-Dmour, R. H. (2021). Factors influencing consumer intentions to adopt e-payment systems: Empirical study. International Journal of Customer Relationship Marketing and Management (IJCRMM), 12(2), 80-99. http://doi.org/10.4018/IJCRMM.2021040105

Banchongduang, S. (2022, December 15). KTB set to roll out digital loans on Pao Tang in 2023. Bangkok Post. Retrieved from https://www.bangkokpost.com/business/general/2460865/ktb-set-to-roll-out-digital-loans-on-pao-tang-in-2023

Bashir, I., & Madhavaiah, C. (2015). Trust, social influence, self-efficacy, perceived risk and internet banking acceptance: An extension of technology acceptance model in Indian context. Metamorphosis, 14(1), 25-38.

Crypto ownership by country. (2023, August 29). Statista. https://www.statista.com/statistics/1202468/global-cryptocurrency-ownership/

Decaro, F., & Saleh, Z.I. (2003). An examination of the internet security and its impact on trust and adoption of online banking.

Digitalisation trends in the Swiss payment landscape. (n.d.). European Payments Council. Retrieved October 2, 2023, from https://www.europeanpaymentscouncil.eu/news-insights/insight/digitalisation-trends-swiss-payment-landscape

Editorial Team. (2022). Krungthai Bank’s super app, ‘Pao Tang’, an all-in-one platform for Thais. INTLBM. https://intlbm.com/2022/07/25/krungthai-banks-super-app-pao-tang-an-all-in-one-platform-for-thais/

Financial Landscape for Digital and Sustainable Economy. (n.d.). Bank of Thailand. Retrieved September 20, 2023, from https://www.bot.or.th/en/financial-innovation/financial-landscape.html

Finma. (2018). FINMA publishes ICO guidelines. Eidgenössische Finanzmarktaufsicht FINMA. Retrieved October 2, 2023, from https://www.finma.ch/en/news/2018/02/20180216-mm-ico-wegleitung/

Gohwong, S. G. (2017). The state of the art and trend of cashless society in Thailand. Asian Political Science Review, 1(2).

IFZ (2023). Crypto Assets Study 2023: An overview of the Swiss and Liechtenstein crypto assets ecosystem. Institute of Financial Services Zug IFZ 2023.

Istrefaj, A. (2023). How can classic payments be linked to Web3 payment systems without acquiring cryptocurrencies? Bachelor Thesis, Lucerne University of Applied Sciences and Arts, 2023.

Khiaonarong, T. (2000). Electronic payment systems development in Thailand. International Journal of Information Management, 20(1), 59-72.

Krungthai Bank. (2022, July 25). Krungthai Bank's super app, 'Pao Tang', an all-in-one platform for Thais. International Business Magazine. Retrieved September 20, 2023, from https://intlbm.com/2022/07/25/krungthai-banks-super-app-pao-tang-an-all-in-one-platform-for-thais/

Krugthai. (2022). Growing together for sustainability: Sustainability Report 2020. Retrieved September 20, 2023, from https://krungthai.com/Download/CSR/CSRDownload_70SD_report_63_en.pdf

Lamsam, A., Pinthong, J., Rittinon, C., Shimnoi, A., & Trakiatikul, P. (2018). The journey to less-cash society: Thailand's payment system at a crossroads. Pouey Ungphakorn Institute for Economic Research, 1–53.

Legal Counsel and Development Department, The Office of the Securities and Exchange Commission. (2018, May). Summary of the Royal Decree on the Digital Asset Businesses B.E. 2561. Retrieved September 8, 2023, from https://www.sec.or.th/EN/Documents/ActandRoyalEnactment/LawReform/summary-decree-digitalasset2561.pdf

Liu, G., Huang, S. P., & Zhu, X. K. (2008, November). User acceptance of Internet banking in an uncertain and risky environment. In 2008 International Conference on Risk Management & Engineering Management (pp. 381-386). IEEE.

Moenjak, T., Kongprajya, & Monchaitrakul, C. (2020). Fintech, financial literacy, and consumer saving and borrowing: The case of Thailand. ADBI Working Paper Series, 110. Retrieved from https://www.adb.org/sites/default/files/publication/575576/adbi-wp1100.pdf

Okanurak, W., Nakarat, W., & Koohasaneh, S. (n.d.). Emergemcu Decree on Digital Asset Businesses B.E.2561(2018). Retrieved September 5, 2023, from https://www.sec.or.th/EN/Documents/EnforcementIntroduction/digitalasset_decree_2561_EN.pdf

Patil, P., Rana, N., Dwivedi, Y., & Abu-Hamour, H. (2018). The role of trust and risk in mobile payments adoption: a meta-analytic review.

Payong Srivanich. (n.d.). Growing together towards sustainability in Thailand. World Finance. Retrieved from https://www.worldfinance.com/banking/growing-together-towards-sustainability-in-thailand

Rankings by Country of Average Monthly Net Salary (After Tax). (n.d.). Retrieved October 2, 2023, from https://www.numbeo.com/cost-of-living/country_price_rankings?itemId=105&displayCurrency=CHF

Schärli, K., Luthiger, R., & Trost, A. (2023). Switzerland - Trends and Developments. In Fintech 2023. MLL Legal. Retrieved from https://mll-legal.com/wp-content/uploads/2023/05/041_SWITZERLAND-TD.pdf

Tamphakdiphanit, J. & Laokulrach, M. (2020). Regulations and Behavioral Intention for Use Cryptocurrency in Thailand. Journal of Applied Economic Sciences, Volume XV, Fall, 3(69) 523-531.

Team, D. A. (2022). Digital adoption in the banking industry. Digital Adoption. Retrieved from https://www.digital-adoption.com/digital-adoption-banking-research/

Whatfix. (n.d.). Digital Adoption & Banking: a 7-Step framework. LinkedIn. Retrieved from https://www.linkedin.com/pulse/digital-adoption-banking-7-step-framework-whatfix/